\documentclass[aps,prd,a4paper,onecolumn,amsmath,showpacs,superscriptaddress,nofootinbib,preprintnumbers,notitlepage]{revtex4-1}
 
\usepackage{verbatim}
\usepackage[T1]{fontenc}
\usepackage[utf8]{inputenc}
\usepackage[american]{babel}
\usepackage{epsfig}
\usepackage{graphicx,subcaption,caption}
\usepackage{pgfplots}
\usepackage{booktabs}
\usepackage{multirow}
\usepackage{dcolumn}
\usepackage{amsmath}
\usepackage{mathtools}
\usepackage{amsfonts}
\usepackage{amssymb}
\usepackage{epstopdf}
\usepackage{bm}
\usepackage{siunitx}
\usepackage{braket}
\usepackage{enumitem}
\usepackage{soul}

\usepackage{transparent}
\usepackage{pifont}
\usepackage[font=small,labelfont=bf]{caption}

\definecolor{navyblue}{rgb}{0.0, 0.0, 0.5}
\definecolor{royalblue}{rgb}{0.25, 0.41, 0.88}
\definecolor{cadmiumgreen}{rgb}{0.0, 0.42, 0.24}
\definecolor{blue-violet}{rgb}{0.54, 0.17, 0.89}
\definecolor{darkviolet}{rgb}{0.58, 0.0, 0.83}
\definecolor{orange(colorwheel)}{rgb}{1.0, 0.5, 0.0}

\usepackage{hyperref}
\hypersetup{
    colorlinks=true, 
    linkcolor=royalblue, 
    citecolor=magenta}

\newcommand\be{\begin{equation}}
\newcommand\ee{\end{equation}}

\newcommand\bea{\begin{eqnarray}}
\newcommand\eea{\end{eqnarray}}

\usepackage{booktabs}
\usepackage{multirow}
\usepackage{dcolumn}
\usepackage{colortbl}

\definecolor{magenta(process)}{rgb}{1.0, 0.0, 0.56}

\definecolor{darkspringgreen}{rgb}{0.09, 0.45, 0.27}

\definecolor{royalblue(web)}{rgb}{0.25, 0.41, 0.88}

\begin{document}

\title{Constant-roll inflation from a fermionic field}

\author{Mehdi Shokri}
\email{mehdishokriphysics@gmail.com}
\affiliation{Department of Physics, Campus of Bijar, University of Kurdistan, Bijar, Iran}

\author{Jafar Sadeghi}
\email{pouriya@ipm.ir}
\affiliation{Department of Physics, University of Mazandaran, P. O. Box 47416-95447, Babolsar, Iran}

\author{Mohammad Reza Setare}
\email{rezakord@ipm.ir}
\affiliation{Department of Physics, Campus of Bijar, University of Kurdistan, Bijar, Iran}

\preprint{}
\begin{abstract}
We study the inflationary period driven by a fermionic field which is non-minimally coupled to gravity in the context of the constant-roll approach. We consider the model for a specific form of coupling and perform the corresponding inflationary analysis. By comparing the result with the Planck observations coming from CMB anisotropies, we find the observational constraints on the parameters space of the model and also the predictions the model. We find that the values of $r$ and $n_{s}$ for $-1.5<\beta\leq-0.9$ are in good agreement with the observations when $|\xi|=0.1$ and $N=60$. \\
\\{\bf PACS:} 98.80.Cq; 98.80.$-$K.
\\{\bf Keywords:} Fermionic field; Constant-roll inflation; Cosmic Microwave Background. 
\end{abstract}
\maketitle
\section{Introduction}
Cosmic inflation has been identified as the most appealing theory removing  shortcomings of the hot big bang model and also the main reliable approach for structure formation of the universe. Furthermore,   primordial gravitational waves are produced through   primordial perturbations and they could be observed, in principle, in the stochastic background \cite{Guth,Sato:1980yn,Kazanas:1980tx,Linde:1981my,Albrecht:1982wi,Lyth:1998xn}. Based on the standard  formalism, a single scalar field (inflaton)  can drive inflation, and then by decaying at the last stage of the period, inflation will be terminated throughout a reheating process \cite{Kofman2,Shtanov}. Several  inflationary models are equipped with  the slow-roll approximation which says inflaton rolls slowly down from the top of potential to the minimum point with a canonical kinetic term. For a long time, the single field inflationary models have been investigated in  literature in detail so that by comparing with the Planck data,  some of them are restricted or even ruled out \cite{martin}. However, some models are still viable with the recent high precision observations \cite{staro,barrow,bezrukov,kallosh1}. One of the most significant properties of these models is that they do not present any non-Gaussianity in their primordial spectrum because of uncorrelated modes of the spectrum \cite{Chen}. In such a case, if the future observations predict non-Gaussianities in the perturbations spectrum, then this class of models will be felt in serious uncertainty. In order to remove the expected obstacles, a new class of inflationary models has been suggested in which inflaton is rolling down with a constant rate during inflation \cite{martin2,Motohashi1,Motohashi2}. This means that $\ddot{\varphi}$ is non-negligible and can be expressed as 
\begin{equation}
\ddot{\varphi}=\beta H\dot{\varphi}
\label{1}
\end{equation}
where $\beta=-(3+\alpha)$ and $\alpha$ is a non-zero parameter. For $\alpha=-3$, the model is reduced to the standard slow-roll. The need to go beyond of the standard slow-roll approximation goes back to the ultra slow-roll regime where the term of $\ddot{\varphi}$ is non-negligible in the Klein-Gordon equation as $\ddot{\varphi}=3H\dot{\varphi}$. This class of inflationary models shows a finite value for the non-decaying mode of curvature perturbations and can be expressed with the familiar form of standard inflation \cite{Inoue}. Although the ultra slow-roll model predicts a large $\eta$ because of the non-negligible term $\ddot{\varphi}$, it is not able to solve the $\eta$ problem introduced in supergravity for the hybrid inflationary models \cite{Kinney}. Moreover, the inflationary solutions of the ultra model are situated in the non-attractor phase of inflation but show a scale-invariance for the scalar perturbation spectrum. Besides, the main problem of ultra models is that the non-Gaussinaity consistency relation of single field models is violated in the presence of ultra condition through super-Hubble evolution of the scalar perturbation \cite{Namjoo}. Despite the above problems, the ultra slow-roll inflationary model is really interesting since it reveals the plenty of unknown and unexplained issues around the inflationary dynamics. Fast-roll models are known as another class of inflationary models introduced in the beyond of the slow-roll approximation in which a fast-rolling stage is considered at the start of inflation and will be connected to the standard slow-roll only after a few e-folds \cite{Contaldi,Lello,Hazra}. This approach can be considered as a correction to slow-roll models in order to improve the consistency with observations.

Constant-roll approach has opened a new window to analyze cosmic inflation due to the consideration of  a constant rate of rolling for inflaton. The model is becoming very popular since a wide range of inflationary models has been investigated in the context of constant-roll inflation \cite{Odintsov,Nojiri,Motohashi9,Cicciarella,Awad,Anguelova,Ito,Yi,Morse,karam1,Ghersi,Lin,Micu,Oliveros,Motohashi3,Kamali,diego,setare,new2}. In the present manuscript, we study a fermionic inflationary model in which a fermionic field which is non-minimally coupled to gravity is the main responsible to drive inflation instead of inflaton introduced in the standard inflationary paradigm. In fact, Noether symmetry arguments of the model show an exponentially expanding universe if a non-minimal coupling between the fermionic field and the Ricci scalar is considered \cite{kramers1}. Fermionic fields also are known as gravitational sources of dark energy as the late-time accelerating of the universe in addition to playing the role of inflaton for the early time acceleration \cite{kramers2}. Moreover, the interaction of a fermionic field with a scalar field can be recognized as the source of interacting dark
matter-dark energy \cite{lepe}. Such theories that deal with  fermionic fields instead of scalar fields are named the Fermions Tensor Theories (FTT) \cite{q1,q2,q3,q4}.

The main aim of this paper is studying the constant-roll inflation for a fermionic inflationary model in which a fermionic field is non-minimally coupled to gravity. We investigate the model for a specific form of coupling and then perform the corresponding inflationary analysis. Finally, we compare the obtained results with the Planck data coming from CMB anisotopies in order to find the observational constraints on the parameters space of the model. The above discussion motivates us to arrange the paper as follows. In \S II, we introduce the properties of the FTT with the corresponding dynamical equations. In \S III, we present the fermionic inflationary model in the context of the constant-roll inflation for the coupling form $f\sim1+\xi\phi$. In \S IV, we analyze the obtained results of the model by comparing with the observational datasets from the Planck and the BICEP2/Keck array satellites. In \S V, we conclude the analysis of the models and draw the possible outlooks.
\section{Fermions Tensor Theory}
Let us start with the inflationary action for a fermionic field $\psi$ non-minimally coupled to to gravity as follows \cite{Bojowald}
\begin{equation}
S=\int{d^{4}x\sqrt{-g}\bigg(\frac{1}{2}f(\phi)R+\frac{i}{2}\big(\bar{\psi}\tilde{\gamma}^{\mu}D_{\mu}\psi-(D_{\mu}\bar{\psi})\tilde{\gamma}^{\mu}\psi\big)}-V(\phi)\bigg)
\label{2}    
\end{equation}
where $g$ is the determinant of the metric $g_{\mu\nu}$, $R=g^{\mu\nu}R_{\mu\nu}$ is the Ricci scalar Ricci scalar, $\psi$ and $\bar{\psi}=\psi^{\dagger}\gamma^{0}$ are the fermionic field and its adjoint field, respectively. $\phi\equiv|\bar{\psi}\psi|$ is the scalar bilinear and $f$ is a function that couples the fermionic field to gravity. Also, $\tilde{\gamma}^{\mu}=e^{\mu}_{\nu}\gamma^{\nu}$ are the generalized gamma matrices where $e^{\mu}_{\nu}$ are the tetrad fields and $D_{\mu}$ is the covariant derivative expressed by 
\begin{equation}
D_{\mu}\psi=\partial_{\mu}\psi-\Omega_{\mu}\psi,\hspace{1cm} D_{\mu}\bar{\psi}=\partial_{\mu}\bar{\psi}+\bar{\psi}\Omega_{\mu}\hspace{1cm}with\hspace{1cm}\Omega_{\mu}=-\frac{1}{4}g_{\rho\sigma}\big(\Gamma^{\rho}_{\mu\delta}-e^{\rho}_{b}(\partial_{\mu}e^{b}_{\delta})\big)\tilde{\gamma}^{\delta}\tilde{\gamma}^{\sigma}
\label{3}   
\end{equation}
where $\Omega_{\mu}$ is the spin connection and $\Gamma^{\mu}_{\nu\delta}$ is the Christoffel symbol. Moreover, $V$ as the fermionic potential can be written in terms of the fermion mass $m$ as follows
\begin{equation}
V=m\bar{\psi}\psi.
\label{4}
\end{equation}
In the following, we consider a spatially flat universe described by a Friedmann-Robertson-Walker (FRW) metric as $ds^{2}=-dt^{2}+a(t)^{2}(dx^{2}+dy^{2}+dz^{2})$
where $t$ and $a$ depict to cosmic time and scale factor, respectively. Therefore, we can state the effective Lagrangian of the model by
\begin{equation}
\mathcal{L}=3a\dot{a}\big(\dot{a}f+a\dot{\phi}f'\big)+\frac{i}{2}a^{3}\big(\dot{\bar{\psi}}\gamma^{0}\psi-\bar{\psi}\gamma^{0}\dot{\psi}\big)+a^{3}V
\label{5}
\end{equation}
where dot and prime refer to the derivative with respect to the cosmic time $t$ and the bilinear field $\phi$, respectively. Using the Euler-Lagrange equation of $\bar{\psi}$ and $\psi$ for the Lagrangian (\ref{5}), we obtain the following equations of motion 
\begin{equation}
\dot{\bar{\psi}}+\frac{3}{2}H\bar{\psi}-iV'_{0}\bar{\psi}\gamma^{0}+6if'\bar{\psi}\gamma^{0}(\dot{H}+2H^{2})=0
\label{6}    
\end{equation}
\begin{equation}
\dot{\psi}+\frac{3}{2}H\psi-iV'_{0}\gamma^{0}\psi+6if'\gamma^{0}\psi(\dot{H}+2H^{2})=0.
\label{7}    
\end{equation}
Multiplying the Eqs. (\ref{6}) and (\ref{7}) by $\psi$ and $\bar{\psi}$, respectively and then adding two equations, the equation of motion of the bilinear field $\phi$ takes the following form
\begin{equation}
\frac{d\phi}{dt}+3H\phi=0   
\label{8}    
\end{equation}
where $H=\dot{a}/a$ denotes the Hubble parameter. Also, by using the Euler-Lagrange equation of $a$ and setting the energy function of the Lagrangian (\ref{5}) as zero, we obtain the acceleration and Friedmann equations, respectively as
\begin{equation}
\frac{\ddot{a}}{a}=-\frac{\rho_{f}+3p_{f}}{6f},\hspace{1cm}H^{2}=\frac{\rho_{f}}{3f}
\label{9}  
\end{equation}
where the energy density $\rho_{f}$ and the pressure $p$ of the fermionic field are given by
\begin{equation}
\rho_{f}=V-3Hf'\dot{\phi},\hspace{1cm}p_{f}=\phi\big(V'-3f'(\dot{H}+2H^{2})\big)-V+\ddot{\phi}f'+2H\dot{\phi}f'+f''\dot{\phi}^{2}.
\label{10}     
\end{equation}
Plugging the above expressions to the Eq. (\ref{9}) and using the the definition of the Hubble constant $H=\dot{a}/a$, the dynamical equations are obtained as
\begin{equation}
3fH^{2}=V-3Hf'\dot{\phi}
\label{11} 
\end{equation}
\begin{equation}
-2f\dot{H}=\phi V'-Hf'\dot{\phi}-3f'\phi(\dot{H}+2H^{2})+\ddot{\phi}f'+f''\dot{\varphi}^{2}
\label{12} 
\end{equation}
where again dot and prime denotes the derivative with respect to $t$ and $\phi$.
\section{Fermionic Constant-roll Inflation}
\begin{figure*}[!hbtp]
	\centering
	\includegraphics[width=.35\textwidth,keepaspectratio]{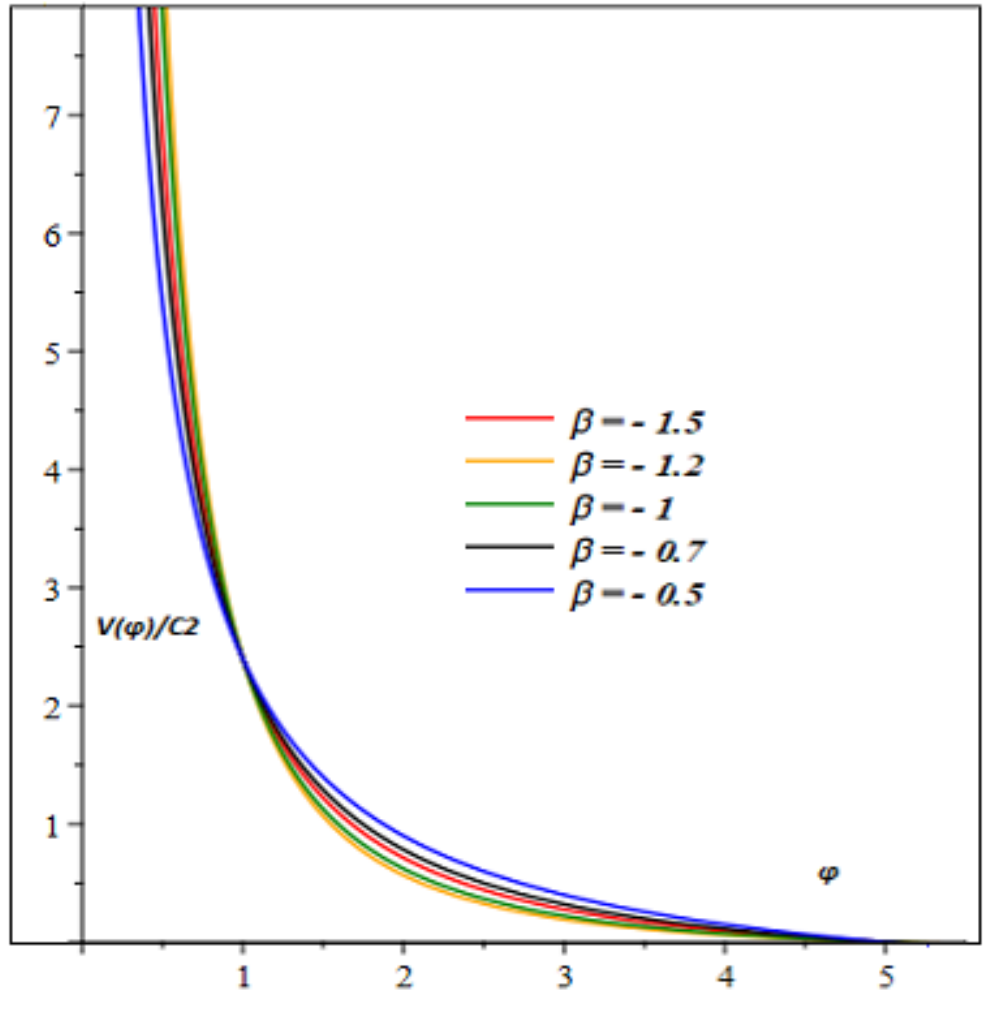}
	\caption{The potential (\ref{16}) plotted versus the bilinear field $\phi$ for different values of $\beta$ when $|\xi|=0.1$.}
	\label{fig1}
\end{figure*}

Now that we know the properties of the FTT, let's study a fermionic inflationary model in the context of the constant-roll idea. By setting the constant-roll condition (\ref{1}) and then using $\dot{H}=\dot{\phi}H'$, the Friedmann equation (\ref{11}) and the equation of motion of the bilinear field (\ref{8}), from (\ref{12}) we have  
\begin{equation}
H(\phi)=C\phi^{-\frac{\beta+3}{3}}
\label{13}
\end{equation}
where $C$ is the integration constant. Also, from the Eq. (\ref{8}), the evolution of the blinear field is given by 
\begin{equation}
\phi(t)=\big(-(\beta+3)Ct\big)^{\frac{3}{3+\beta}}.
\label{14}
\end{equation}
Since Noether symmetry discussions of the model show an exponential expansion for the universe if $f'\neq0$ \cite{kramers1,kramers2}, we consider the coupling form 
\begin{equation}
f=1+\xi\phi
\label{15}    
\end{equation}
where $\xi$ is the coupling constant. From the Eq. (\ref{11}), we obtain the corresponding potential as
\begin{equation}
V(\phi)=3C^{2}(1-2\xi\phi)\phi^{-\frac{2(\beta+3)}{3}} 
\label{16}
\end{equation}
which $\phi_{critical}=\frac{1}{2\xi}$ is the barrier that the bilinear field can not cross it. In Fig. \ref{fig1}, we present the behaviour of the above potential for different values of $\beta$ when $|\xi|=0.1$ in which the bilinear field $\phi$ sharply rolls down from top of the potential to the minimum point. Now, by using the conformal transformation $\hat{g}_{\mu\nu}=\Omega^{2} g_{\mu\nu}$ where the conformal factor $\Omega=\Omega(\phi(x))$ is a non-null and differentiable  function, we can move from the Jordan frame to the Einstein frame with the conformal factor $\Omega^{2}=f$. Therefore, the action in the Einstein frame is given by
\begin{equation}
S_{E}=\int
d^{4}x\sqrt{-\hat{g}}\bigg(\frac{\hat{R}}{2}-\frac{i}{2}F^{2}\big(\hat{\bar{\psi}}\tilde{\gamma}^{\mu}\hat{D}_{\mu}\hat{\psi}-(\hat{D}_{\mu}\hat{\bar{\psi}})\tilde{\gamma}^{\mu}\hat{\psi}\big)
-\hat{V}(\hat{\phi})\bigg).
\label{17}
\end{equation}
In the new frame, we deal with a redefined bilinear field $\hat{\phi}$ and its potential 
\begin{equation}
F^{2}\equiv\bigg(\frac{d\hat{\phi}}{d\phi}\bigg)^{2}=\frac{2f+3f'^{2}}{2f^{2}},\quad\quad\hat{V}(\hat{\phi})\equiv\frac{V(\phi)}{\Omega^{4}}=\frac{V(\phi)}{f^{2}}.
\label{18}
\end{equation}
Now, we can calculate the slow-roll parameters of the model
\begin{equation}
\hat{\epsilon}\equiv\frac{1}{2}\bigg(\frac{\hat{V}'}{\hat{V}}\bigg)^{2}=\frac{4\Big(\phi^{2}(2\beta+9)\xi^{2}+\phi(\beta-3)\xi-(\beta+3)\Big)^{2}}{9\phi^{2}\big(2\phi\xi+3\xi^{2}+2\big)(2\phi\xi-1)^{2}},
\label{19}
\end{equation}
\begin{eqnarray}
&\!&\!\hat{\eta}\equiv\frac{\hat{V}''}{\hat{V}}=\frac{1}{9\phi^{2}\big(2\phi\xi+3\xi^{2}+2\big)^{2}(2\phi\xi-1)}\bigg\{12(2\beta+9)^{2}\phi^{3}\xi^{5}+4\Big((8\beta^{2}+78\beta+189)\phi^{2}+18\beta^{2}+90\beta-27\Big)\phi^{2}\xi^{4}+\nonumber\\&\!&\!
+\Big((80\beta^{2}+564\beta+648)\phi^{3}-(252\beta+756)\phi\Big)\xi^{3}+12(2\beta+9)\Big(2(\beta-3)\phi^{2}-(\beta+3)\Big)\xi^{2}-4(\beta+3)\phi(4\beta+63)\xi-\nonumber\\&\!&\!
-(16\beta^{2}+120\beta+216)\bigg\},
\label{20}
\end{eqnarray}
\begin{eqnarray}
&\!&\!\hat{\zeta}^{2}\equiv\frac{\hat{V}'\hat{V}'''}{\hat{V}^{2}}=\frac{16\Big(\phi^{2}(2\beta+9)\xi^{2}+\phi(\beta-3)\xi-(\beta+3)\Big)}{567\phi^{4}\big(2\phi\xi+3\xi^{2}+2\big)^{4}(2\phi\xi-1)^{2}}\Bigg\{63(2\beta+9)^{3}\phi^{4}\xi^{8}+21\Big(2(8\beta+33)(\beta+6)(2\beta+9)\phi^{2}+\nonumber\\&\!&\!
+60\beta^{3}+648\beta^{2}+1836\beta+405\Big)\phi^{3}\xi^{7}+\Big(28(\beta+6)(2\beta+9)(4\beta+21)\phi^{4}+21(\beta+6)\big(112\beta^{2}
+714\beta+675\big)\phi^{2}+\nonumber\\&\!&\!
+189(4\beta^{2}+6\beta-111)(\beta+3)\Big)\phi^{2}\xi^{6}+63\Big(2(\beta+6)\big(8\beta^{2}+62\beta+87\big)\phi^{4}+\big(\frac{128}{3}\beta^{3}+378\beta^{2}+471\beta-1350\big)\phi^{2}-2\nonumber\\&\!&\!
\times(2\beta+9)(2\beta+39)(\beta+3)\Big)\phi\xi^{5}+7\Big(8\big(333\beta^{2}+891\beta+30\beta^{3}\big)\phi^{4}+(32\beta^{2}-330\beta-2997)(\beta+3)\phi^{2}-18(\beta+6)\nonumber\\&\!&\!
\times(2\beta+9)(\beta+3)\Big)\xi^{4}+7\Big(2(80\beta^{3}+468\beta^{2}-1422\beta-7371)\phi^{2}-3(2\beta+9)(\beta+3)(16\beta+213)\Big)\phi\xi^{3}-168\nonumber\\&\!&\!
\times\Big(\frac{3}{2}(26\beta+147)\phi^{2}+2\beta^{2}+21\beta+54\Big)(\beta+3)\xi^{2}-42(4\beta+45)(2\beta+9)(\beta+3)\phi\xi-56(\beta+6)(2\beta+9)(\beta+3)\Bigg\}
\label{21}
\end{eqnarray}
where prime implies to the derivative with respect to the bilinear field $\phi$ and inflation ends when the condition $\hat{\epsilon} = 1$ or $\hat{\eta} = 1$ is fulfilled. The number of e-folds is given by
\begin{equation}
\hat{N}\equiv\int^{\hat{\phi}_{i}}_{\hat{\phi}_{f}}{\frac{1}{\sqrt{2\hat{\epsilon}}}}d\hat{\phi}
\label{22}
\end{equation}
and also the spectral parameters, \textit{i.e.} the spectral index, its running and the tensor-to-scalar ratio are defined by 
\begin{equation}
\hat{n}_{s}=1-6\hat{\epsilon}+2\hat{\eta},\quad\quad\quad
\hat{\alpha}_{s}=16\hat{\epsilon}{\hat{\eta}}-24\hat{\epsilon}^{2}-2\hat{\zeta}^{2},\quad\quad\quad\hat{r}=16\hat{\epsilon}.
\label{33}  
\end{equation}
\begin{figure*}[!hbtp]
	\centering
	\includegraphics[width=.60\textwidth,keepaspectratio]{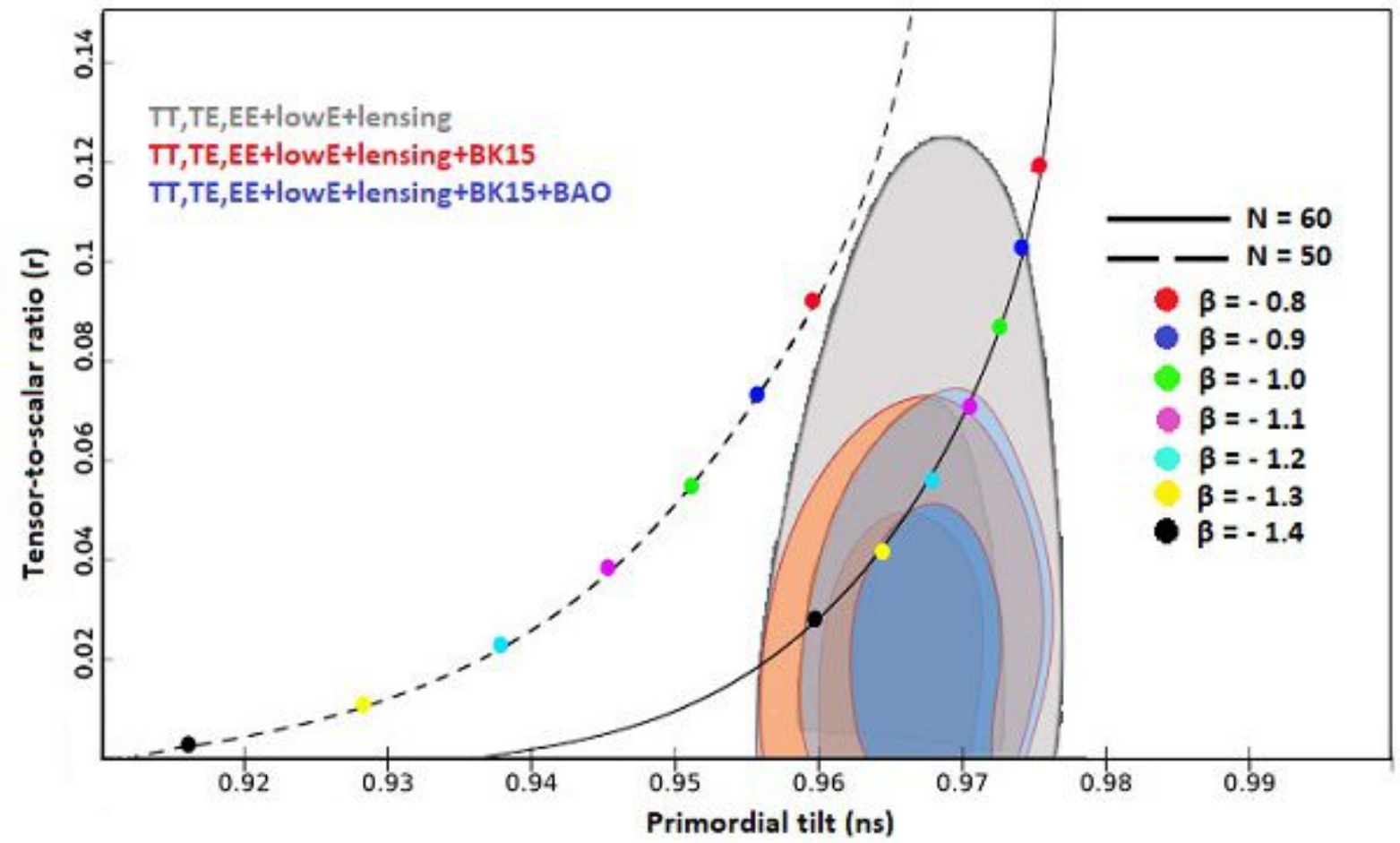}
	\caption{The marginalized joint 68\% and 95\% CL regions for $n_{s}$ and $r$ at $k = 0.002$ Mpc$^{-1}$ from  Planck alone and in combination with BK15 or BK15+BAO data \cite{cmb} and the $n_{s}-r$ constraints on the model with the coupling (\ref{15}). The dashed and solid lines represent $N=50$ and $N=60$, respectively. The results are obtained for different values of $\beta$ when $|\xi|=0.1$.}
	\label{fig2}
\end{figure*}
\section{Constraints from CMB Anisotropies}
Let's compare the obtained results with the observational datasets coming from CMB anisotropies. In Figure \ref{fig2}, we present the $n_{s} - r$ constraints coming from the marginalized joint 68\% and 95\% CL regions of the Planck 2018 alone and in combination with BK15 or BK15+BAO data on the fermionic inflationary model with the coupling (\ref{15}) introduced in the context of the constant-roll approach. The figure is drawn for different values of $\beta$ in the cases $N=50$ (dashed line) and $N=60$ (solid line) when $|\xi|=0.1$. By considering the CMB anisotropies datasets from the Planck alone, we find that all values of $\beta$ in the case of $N=50$ show a reasonable tensor-to-scalar ratio $r$ below the upper limit while the corresponding values of the spectral index $n_{s}$ are not in good agreement with observations. The situation in the case of $N=60$ is much better than $N=50$. As we can see, the cases of $\beta>-0.9$ and $\beta\leq-1.5$ are excluded since the obtained values of $r$ and $n_{s}$ are out of the observational region. The cases of $-1.15\leq\beta\leq-0.9$ and $-1.4<\beta<-1.15$ show observationally acceptable values of $r$ and $n_{s}$ at the 68\% and 95\% CL, respectively. By combination of the BK15 data and the Planck data, the situation of $N=50$ is the similar to the previous observational case and all values of $\beta$ provide disfavored values of $r$ and $n_{s}$. For $N=60$ the cases of $\beta\geq-1.1$ and $\beta\leq-1.5$ show undesirable values of $r$ and $n_{s}$ while the cases of $-1.25\leq\beta<-1.1$ and $-1.4<\beta<-1.25$ present the values compatible with the observations at the 68\% and $95\%$ CL, respectively. As a full consideration, we compare the obtained results with the observations coming from the Planck in combination with BK15+BAO data. In such a case, still $N=50$ is fully excluded by the observations. Also, $\beta>-1.1$ and  $\beta<-1.3$ are not in good agreement with data while the cases of $-1.25\leq\beta\leq-1.1$ and $-1.3\leq\beta<-1.25$ show desirable values of $r$ and $n_{s}$ at the 68\% and 95\% CL, respectively. 
\section{Conclusion}
In this paper, we have studied the constant-roll inflation driven by a fermionic field instead of inflaton which is non-minimally coupled to gravity. First, we have introduced the Fermions Tensors Theories (FTT) briefly and then have investigated the constant-roll condition for the model with the coupling form $f\sim1+\xi\varphi$ which has been approved by the Noether symmetry. At the next step, we calculated the spectral parameters of the model and finally have compared the results with the CMB observations in order to find the observational constraints on the parameters space, in particular, the constant-roll parameter $\beta$. We have found that the obtained values of $r$ and $n_{s}$ for $-1.5<\beta\leq-0.9$ are fully compatible with the CMB anisotriopes observations when $|\xi|=0.1$ and $N=60$.
\bibliographystyle{ieeetr}
\bibliography{biblo}
\end{document}